\documentclass[fp,twocolumn]{jpsj3}
\usepackage{txfonts}
\usepackage{bm}
\usepackage[dvipdfmx]{graphicx}
\usepackage[dvipdfmx]{color}
\usepackage{amsmath, amssymb}
\usepackage{braket}
\usepackage{color}
\usepackage{comment}

\title{Machine-Learning Detection of the Berezinskii-Kosterlitz-Thouless Transitions}
\author{Masahito Mochizuki\thanks{masa{\_}mochizuki@waseda.jp}, and Yusuke Miyajima}
\inst{Department of Applied Physics, Waseda University, Okubo, Shinjuku-ku, Tokyo 169-8555, Japan}

\abst{The Berezinskii-Kosterlitz-Thouless (BKT) transition is a typical topological phase transition defined between binding and unbinding states of vortices and antivortices, which is not accompanied by spontaneous symmetry breaking. It is known that the BKT transition is difficult to detect from thermodynamic quantities such as specific heat and magnetic susceptibility because of the absence of anomaly in free energy and significant finite-size effects. Therefore, methods based on statistical mechanics which are commonly used to discuss phase transitions cannot be easily applied to the BKT transition. In recent years, several attempts to detect the BKT transition using machine-learning methods based on image recognition techniques have been reported. However, it has turned out that the detection is difficult even for machine learning methods because of the absence of trivial order parameters and symmetry breaking. Most of the methods proposed so far require prior knowledge about the models and/or preprocessing of input data for feature engineering, which is problematic in terms of the general applicability. In this article, we introduce recent development of the machine-learning methods to detect the BKT transitions in several spin models. Specifically, we demonstrate the success of two new methods named temperature-identification method and phase-classification method for detecting the BKT transitions in the $q$-state clock model and the XXZ model. This progress is expected to sublimate the machine-learning-based study of spin models for exploring new physics beyond simple benchmark test.}

\begin{document}
\maketitle
\section{Introduction}
Classical spin models such as Ising models, clock models, XY models, and XXZ models are of interest not only as theoretical models of magnetic materials but also as fundamental mathematical models for various natural and social phenomena and information processing architectures~\cite{Cuevas16,Stengele23}. Their statistical mechanical properties have been intensively studied for many years. In particular, the phase-transition phenomena exhibited by these spin models are important topics in statistical mechanics and condensed-matter physics, which have been investigated intensively not only for pure theoretical science but also for exploration of material functions of magnets.

\begin{figure} [tb]
\includegraphics[scale=0.5]{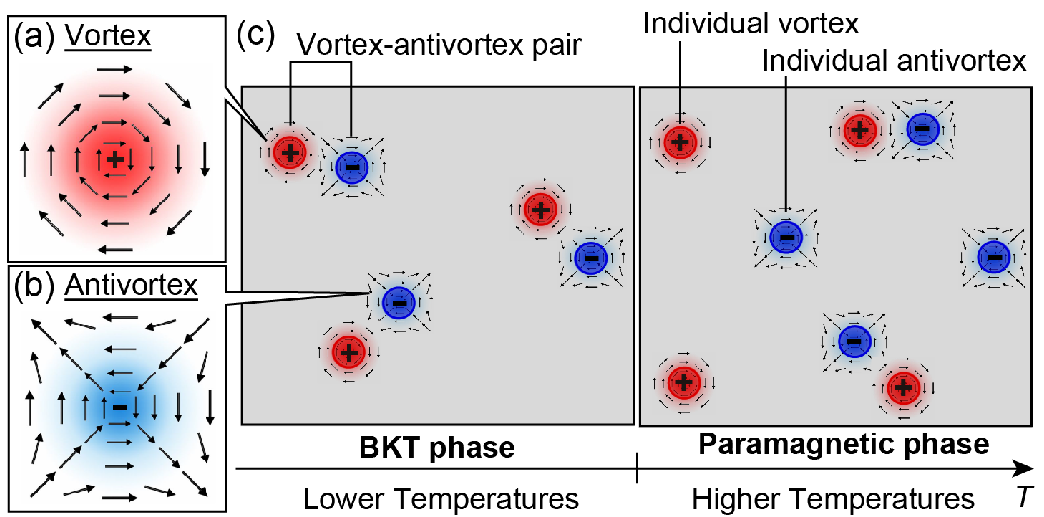}
\caption{(Color online) (a),(b) Schematics of a vortex and an antivortex formed by spin vectors in classical lattice spin models. (c) Schematic phase diagram of the BKT transition. With decreasing temperature, a phase transition occurs from the high-temperature paramagnetic phase to the low-temperature BKT phase. There appear individual vortices and antivortices in the paramagnetic phase, whereas they always appear as vortex-antivortex pairs in the BKT phase. The BKT transition is a topological phase transition, which is not accompanied by spontaneous symmetry breaking.}
\label{Fig01}
\end{figure}

Some of these spin models exhibit topologically characterized phase transitions with no symmetry breaking in addition to ordinary phase transitions accompanied by symmetry breaking. One example is the Berezinskii-Kosterlitz-Thouless (BKT) transition~\cite{Berezinskii71,Berezinskii72,Kosterlitz73,Kosterlitz74}, which shows up in some two-dimensional classical spin models, e.g., XY models, XXZ models, and the $q$-state clock models. In this article, we discuss recently proposed powerful and efficient machine learning methods to detect this phase transition~\cite{Miyajima21,Miyajima23}. The BKT transition can be regarded as a vortex-antivortex bound-unbound phase transition, which occurs between the paramagnetic phase at high temperatures and the BKT phase at low temperatures [see Figs.~\ref{Fig01}(a)-(c)]. In the BKT phase, vortices and antivortices formed by spin vectors appear always in pairs. On the contrary, they appear in individual form in the paramagnetic phase. According to the Mermin-Wagner theorem, long-range order with symmetry breaking is forbidden at finite temperatures for models with continuous spin variables such as the XY models and the classical Heisenberg models in one and two dimensions~\cite{Mermin66}. Hence, the BKT transition is an important phase transition in low-dimensional spin systems. Its research has a history of over 50 years since the first prediction by Berezinskii~\cite{Berezinskii71,Berezinskii72} and the subsequent theoretical confirmation by Kosterlitz and Thouless~\cite{Kosterlitz73, Kosterlitz74}.

%%%%%%%%%%%%%%%%%%%%%%%%%%%%%%%%%%%%%%%%%%%%%%%%%%%
%%%%%%%%%%%%%%%%%%%%%%%%%%%%%%%%%%%%%%%%%%%%%%%%%%%
However, the BKT transition is known to be difficult to treat by numerical analyses based on statistical mechanics such as Monte Carlo methods as compared to ordinary phase transitions. The reason is that thermodynamic quantities such as specific heat and magnetic susceptibility, which are usually used for theoretical detection of ordinary phase transitions, cannot be easily used to determine the BKT-transition point. For example, in the case of ordinary phase transitions, specific heat shows an anomaly such as peak or jump at the transition point, but in the case of the BKT transition, an anomaly of the specific heat does not correspond to its transition point. On the other hand, it is, in principle, possible to determine the BKT-transition point from the anomaly of magnetic susceptibility. However, because of significant finite-size effects with logarithmic correction terms, extrapolation to the thermodynamic limit based on the conventional finite-size scaling is difficult.

Under these circumstances, several attempts have been made to detect the BKT transitions by using machine learning methods rather than conventional methods based on statistical mechanics recently. However, it is not necessarily easy to detect the BKT transitions even for machine learning methods. This is because the methods are basically based on image recognition techniques~\cite{Ohtsuki20}, but the BKT transition does not have any characteristic order parameters and symmetry breaking. In fact, there are only limited examples of successful detections of the BKT transitions by machine learning~\cite{Miyajima21,Miyajima23,Richter-Laskowska18,Beach18,ZhangW19,Rodriguez-Nieva19,Shiina20,Tomita20,TranQH21,WangJ21,Mendes-Santos21,Singh21,Haldar22}. Moreover, it should be mentioned that most of the successful methods are supervised methods, while only few unsupervised methods succeeded in the detection. One successful unsupervised method is based on dimensionality reduction using a diffusion-map technique~\cite{WangJ21}. This situation is in sharp contrast to the case of ordinary phase transitions with symmetry breaking in, e.g.,  Ising models and Potts models, to which the machine learning methods can easily detect their phase transitions because there are clear changes in the order parameters through the phase transitions. For the ordinary phase transitions, not only supervised methods but also many unsupervised methods have successfully been used for their detections~\cite{Tanaka17,Carrasquilla17,Arai18,Suchsland18,Hu17,Wetzel17,WangL16,Ponte17,Nieuwenburg17,Liu18,Giannetti19,Bachtis20a}.

Furthermore, most of the methods which succeeded in detection of the BKT transitions, in fact, require preprocessed data rather than raw spin-configuration data for input data. For example, spatial configurations of vortices, histograms of spin orientations, or spin correlation functions are required to be prepared from the raw spin-configuration data in advance. In other words, these methods depend on arbitrary selection of feature quantities of the transitions and phases and thus are not applicable to general cases. A more critical problem is that these methods require prior knowledge about the model in advance such as the number of phases and approximate transition temperatures. As can be easily recognized, this is a contradictory requirement in the sense that we have to know properties and behaviors of the model to be studied before the investigation, which can hardly withstand exploration of new physics. It is difficult to detect the BKT transitions with neither symmetry breaking nor well-defined order parameters for machine learning methods based on the pattern classification or image recognition techniques. The establishment of powerful and versatile machine learning methods for the BKT transitions has been highly demanded.

In the meanwhile, there are no reports of experimental observation of BKT transition in real magnetic materials, despite a half-century of long research history~\cite{Cuccoli03a,Cuccoli03b,HuZ20,Tutsch14,Kohama11}. One of its reasons is that we do not have a general theoretical framework to discuss a possibility of the BKT transition in complex spin models describing real materials. In other words, we do not have means to theoretically search for materials that exhibit the BKT transition using microscopic models which contain several complex interactions (e.g., Dzyaloshinskii-Moriya interactions and biquadratic interactions) and several magnetic anisotropies in addition to ordinary exchange interactions. To further deepen and develop the research on topological phase transitions, experimental studies using real magnetic materials that exhibit BKT transitions are essentially important. For this purpose, construction of machine learning methods to detect topological phases and topological phase transitions in microscopic spin models for real magnetic materials is an urgent issue.

In this article, we present recent attempts to construct machine learning methods aiming at the detection of BKT transitions in spin models. In particular, we focus on recently proposed two methods, which are named phase-classification method and temperature-identification method. We demonstrate their efficiency by applying these methods to two important spin models that exhibit BKT transitions, i.e., the $q$-state clock models and the XXZ models. It is shown that the methods can detect the BKT transitions and determined their transition temperatures in most cases with high accuracy where only minimal prior knowledge about the models and prior data-processing for feature engineering are required. The rest of this article is structured as follows. We introduce properties of the BKT transition in in Sect.2 and several classical spin models which exhibit the BKT transitions in Sect.3. In Set.4, we explain recently proposed powerful machine learning methods to detect the BKT transitions. In Sects.5-7, we demonstrate that the methods can detect both the BKT transitions and the second-order phase transitions in the $q$-state clock models and the XXZ models on square lattices. In Sect.8, we compare these new machine learning methods with conventional methods based on statistical mechanics such as the Monte Carlo methods as well as previously proposed other machine learning methods by particularly focusing on the advantages and disadvantages. Section 9 is devoted to the summary and conclusion.

\section{Berezinskii-Kosterlitz-Thouless transition}
In 1971, Berezinskii argued that the spatial dependence of spin correlation function differs between low and high temperatures in the XY model~\cite{Berezinskii71,Berezinskii72}. On this basis, he predicted the presence of a novel type of phase transition in this model. Subsequently, Kosterlitz and Thouless confirmed presence of the predicted phase transition and elucidated its physical properties using renormalization group analyses~\cite{Kosterlitz73, Kosterlitz74}. This phase transition is called Berezinskii-Kosterlitz-Thouless (BKT) transition after the names of these three physicists. The BKT transition is a kind of topological phase transition beyond the framework of Landau theory based on order parameters. The BKT transition is defined as a phase transition between bounded and unbounded states of vortices and antivortices and does not exhibit symmetry breaking. Here vortices and antivortices in classical spin models are formed by spin vectors. In the case of a square lattice, as we trace the four corner sites of each plaquette in a counterclockwise manner, the (anti)vortex is defined as a spin configuration at the four sites rotating in a counterclockwise (clockwise) sense. At low temperatures, vortices and antivortices always appear forming vortex-antivortex pairs, and this state is called BKT phase. On the contrary, a paramagnetic phase appears at high temperatures, in which there exist many individual or unbounded vortices and antivortices. The BKT phase exhibits behaviors distinct from ordinary phases with broken symmetry. For example, the spin correlation length always of power-law decay with respect to distance.

It is known that the BKT transition is difficult to detect by computational methods based on statistical mechanics because of the absence of symmetry breaking, absence of anomaly of the free energy at the transition point, and significant finite-size effects with logarithmic correction terms. In particular, it is difficult to apply methods usually employed to study ordinary phase transitions. For example, uniform magnetization is zero at all temperatures in the thermodynamic limit and cannot be exploited as an order parameter or a signal of the BKT transition. Since the free energy does not have anomaly with respect to temperature, the specific heat shows no anomaly at the BKT-transition point. On the other hand, the magnetic susceptibility shows divergence or jump at the BKT-transition point, and thus might be used to determine the transition temperature in principle. However, finite-size effects are significant, which contain logarithmic correction terms, and thus the determination of BKT-transition temperature requires an enormous computational cost.

A physical quantity called helicity modulus $\gamma$ is often used to detect the BKT transition and to identify its transition temperature by methods based on statistical mechanics~\cite{Fisher73}. This quantity describes hardness of spin order and is defined as the second-order response coefficient of free energy with respect to the global twist of the spin alignment. Its definition is given by,
%%%%%%%%%%%%%%%%%%%%%%%%%%%%%%%%%%%%%%%%%%%%%%%%%%%
\begin{align}
F(\delta)-F(0)=\frac{\gamma}{2}\delta^2 + \mathcal{O}(\delta^4),
\end{align}
%%%%%%%%%%%%%%%%%%%%%%%%%%%%%%%%%%%%%%%%%%%%%%%%%%%
where $F$ is the free energy. Here it is assumed that the rightmost spin in the system is twisted by an infinitesimal angle $\delta$ relative to the leftmost spin. More specifically, we consider the situation that the spins are gradually twisted with an equivalent twisting angle of $\delta/L$ from the left end to the right end of the system where the lattice size is $L$ in one-dimensional directions. 

In the thermodynamic limit, the helicity modulus exhibits a discontinuous jump from $0$ to $ 2T_{\rm BKT}/\pi$ at the BKT-transition temperature $T_{\rm BKT}$~\cite{Nelson77,Weber88}. Although the jump becomes obscure in finite-size systems, the transition temperature in the thermodynamic limit can be extrapolated through finite-size scaling from tentative transition temperatures for several system sizes, which are evaluated from intersection of the temperature-dependence plot of helicity modulus and the line of $2T/ \pi$. However, the helicity modulus has some problems for practical use. First, it cannot be applied to spin systems with discrete degrees of freedom. Second, unlike specific heat and magnetic susceptibility, there is no universal expression which can be used to calculate its thermal average for any spin models. When performing numerical calculations, it is necessary to derive the model-dependent expression, which is not suitable for systematic detection. It becomes difficult especially for complex spin models which include magnetic anisotropies and higher-order interactions.

\section{Spin models}
The $XY$ model, the $q$-state clock model, and the classical XXZ model are typical examples of classical spin models which exhibit the BKT transition. We explain each of these models in the following.

\subsection{XY models}
The Hamiltonian of the $XY$ model is given by,
%%%%%%%%%%%%%%%%%%%%%%%%%%%%%%%%%%%%%%%%%%%%%%%%%%%
\begin{align}
\mathcal{H} = -J \sum_{\langle i,j \rangle} \bm{S}_i \cdot \bm{S}_j
=-J \sum_{\langle i,j \rangle} \cos(\phi_i - \phi_j),
\end{align}
%%%%%%%%%%%%%%%%%%%%%%%%%%%%%%%%%%%%%%%%%%%%%%%%%%%
with
%%%%%%%%%%%%%%%%%%%%%%%%%%%%%%%%%%%%%%%%%%%%%%%%%%%
\begin{align}
\bm{S}_i=\left(\cos\phi_i, \sin\phi_i\right).
\end{align}
%%%%%%%%%%%%%%%%%%%%%%%%%%%%%%%%%%%%%%%%%%%%%%%%%%%
Normalized classical spin vectors $\bm{S}_i$ ($|\bm{S}_i|$=1) on a square lattice are ferromagnetically interacting via the nearest-neighbor exchange interactions $J(>0)$, which can be oriented only within the two-dimensional plane. Here the sum over $\langle i,j \rangle$ is meant to be taken over pairs of neighboring spins. 

\subsection{$q$-state clock model}
%%%%%%%%%%%%%%%%%%%%%%%%%%%%%%%%%%%%%%%%%%%%%%%%%%%
\begin{figure} [tb]
\includegraphics[scale=1.0]{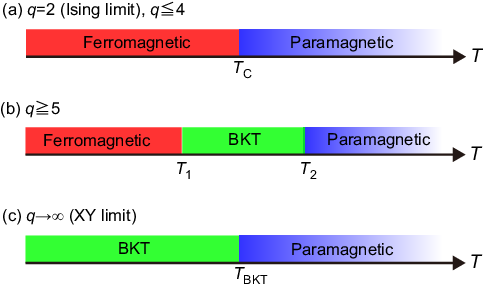}
\caption{(Color online) Schematic temperature phase diagrams of the $q$-state clock model on a square lattice. (a) When $2 \le q \le 4$, the model exhibits a single second-order phase transition from paramagnetic to ferromagnetic phases at $T_{\rm c}$. (b) When $q \ge 5$, the model exhibits successive phase transitions from paramagnetic to BKT to ferromagnetic phases at $T_2$ and $T_1$. (c) In the limit of $q \rightarrow \infty$, the model is equivalent to the XY model, which exhibits a single BKT transition from paramagnetic to BKT phases at $T_{\rm BKT}$.}
\label{Fig02}
\end{figure}
%%%%%%%%%%%%%%%%%%%%%%%%%%%%%%%%%%%%%%%%%%%%%%%%%%%
The Hamiltonian of the $q$-state clock model is given by,
%%%%%%%%%%%%%%%%%%%%%%%%%%%%%%%%%%%%%%%%%%%%%%%%%%%
\begin{align}
\mathcal{H} = -J \sum_{\langle i,j \rangle} \bm{S}_i \cdot \bm{S}_j
\end{align}
%%%%%%%%%%%%%%%%%%%%%%%%%%%%%%%%%%%%%%%%%%%%%%%%%%%
with
%%%%%%%%%%%%%%%%%%%%%%%%%%%%%%%%%%%%%%%%%%%%%%%%%%%
\begin{align}
\bm{S}_i = \left( \cos \frac{2\pi k}{q}, \sin \frac{2\pi k}{q} \right)  \hspace{1cm} (k=0, 1, \cdots q-1)
\end{align}
%%%%%%%%%%%%%%%%%%%%%%%%%%%%%%%%%%%%%%%%%%%%%%%%%%%
Normalized classical spin vectors $\bm S_i$ ($|\bm S_i|$=1) are considered on a square lattice, which are ferromagnetically interacting via the nearest-neighbor exchange interactions $J(>0)$. The spin vectors can be oriented only in the $q$ discretized directions within the two-dimensional plane. 

In the discrete limit of $q=2$, this model becomes identical to the Ising model, which shows a single second-order phase transition from paramagnetic to ferromagnetic phases with decreasing temperature. On the other hand, in the continuous limit of $q \rightarrow \infty$, this model becomes equivalent to the XY model, which shows a single BKT transition with decreasing temperature. The $q$-state clock model is an important model which connects these two representative models and thus has been studied intensively. According to previous studies, this model is known to exhibit distinct behaviors depending on the number of states $q$ [see Figs.\ref{Fig02}(a)-(c)]~\cite{Jose78,Elitzur79,Tobochnik82,Kumano13}. When $q \leq 4$, this model exhibits a single second-order phase transition similar to the Ising model [Fig.\ref{Fig02}(a)]. On the other hand, when $q \geq 5$, this model exhibits successive two phase transitions associated with the BKT phase [Fig.\ref{Fig02}(b)], that is, the paramagnetic phase, BKT phase, and ferromagnetic phase successively appear with decreasing temperature. In the XY limit of $q \rightarrow \infty$, the lowest-temperature ferromagnetic phase disappears, resulting in the single BKT transition [Fig.\ref{Fig02}(c)]. Note that the helicity modulus cannot be exploited to detect the BKT transition for this model because of the discretized spin degree of freedom. In the following, we discuss the cases with $q=4$ and $q=8$ as typical examples of these two cases, respectively.

\subsection{XXZ model}
%%%%%%%%%%%%%%%%%%%%%%%%%%%%%%%%%%%%%%%%%%%%%%%%%%%
\begin{figure} [tb]
\includegraphics[scale=1.0]{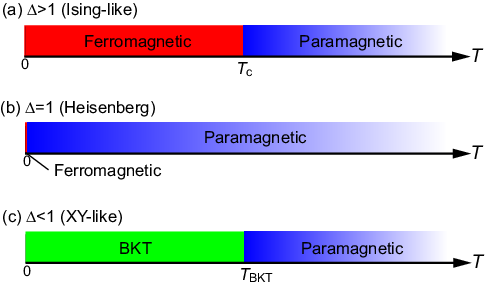}
\caption{(Color online) Schematic temperature phase diagrams of the XXZ model on a square lattice. (a) For the Ising-like case with $\Delta>1$, the model exhibits a single second-order phase transition from paramagnetic to ferromagnetic phases at $T_{\rm c}$. (b) When $\Delta=1$, the model becomes identical to the classical Heisenberg model which does not exhibit any phase transition at finite temperature. (c) For the XY-like case with $\Delta<1$, the model exhibits a single BKT transition from paramagnetic to BKT phases at $T_{\rm BKT}$.}
%(d) Temperature dependence of the helicity modulus $\gamma$ for a system of $L \times L$ sites with $L$=32 calculated using the Monte Carlo method for various values of $\Delta$. This quantity becomes finite below a certain temperature in the XY-like case with $\Delta<1$, whereas it always nearly equals to zero in the Ising-like case with $\Delta>1$.
\label{Fig03}
\end{figure}
%%%%%%%%%%%%%%%%%%%%%%%%%%%%%%%%%%%%%%%%%%%%%%%%%%%
The Hamiltonian of the classical XXZ model is given by,
%%%%%%%%%%%%%%%%%%%%%%%%%%%%%%%%%%%%%%%%%%%%%%%%%%%
\begin{align}
\mathcal{H} = -J \sum_{\langle i,j \rangle} \left(S_i^x S_j^x + S_i^y S_j^y + \Delta S_i^z S_j^z \right)
\end{align}
%%%%%%%%%%%%%%%%%%%%%%%%%%%%%%%%%%%%%%%%%%%%%%%%%%%
with
%%%%%%%%%%%%%%%%%%%%%%%%%%%%%%%%%%%%%%%%%%%%%%%%%%%
\begin{align}
\bm{S}_i = \left( \sqrt{1-(S_i^z)^2}\cos\phi_i, \sqrt{1-(S_i^z)^2}\sin\phi_i, S_i^z \right)
\end{align}
%%%%%%%%%%%%%%%%%%%%%%%%%%%%%%%%%%%%%%%%%%%%%%%%%%%
Normalized three-dimensional classical spin vectors $\bm S_i$ ($|\bm S_i|$=1) are considered on a square lattice, which are ferromagnetically interacting via the nearest-neighbor exchange interactions with $J(>0)$ and $\Delta>0$. Here $\Delta$ is the anisotropy parameter of the exchange interaction in the spin space. The XXZ model has been intensively studied as a model for superconducting and magnetic thin films~\cite{Hikami80,Kawabata86,Serena93,Cuccoli95,Pires96,Lee05,Aoyama19,Shirinyan19}. Temperature phase diagrams of this model are shown in Figs.~\ref{Fig03}(a)-(c). Behaviors of this model depend on $\Delta$. When $\Delta>1$, the model exhibits a single second-order phase transition from the paramagnetic phase to the ferromagnetic phase with decreasing temperature, similar to the Ising model. On the other hand, when $0<\Delta<1$, the model exhibits a BKT transition from the paramagnetic phase to the BKT phase, similar to the XY model. When $\Delta=1$, the model is identical to the two-dimensional classical Heisenberg model, which does not show any phase transition at finite temperatures and the paramagnetic phase is always observed in accordance with the Mermin-Wagner theorem~\cite{Mermin66}. In the following, we discuss the cases with $\Delta$=1.05 and $\Delta$=0.95 as typical examples of the Ising-like case and the XY-like case, respectively.
\section{Machine Learning Methods}
In this section, we introduce two recently proposed versatile machine learning architectures to detect phase transitions, that is, the phase-classification (PC) method and the temperature-identification (TI) method. These neural networks can be easily implemented by the machine learning library KERAS~\cite{KERAS15}.

\subsection{Phase-classification (PC) method}
%%%%%%%%%%%%% Fig4 %%%%%%%%%%%%%%%%%%%%%%%
\begin{figure}[tb]
\includegraphics[scale=1.0]{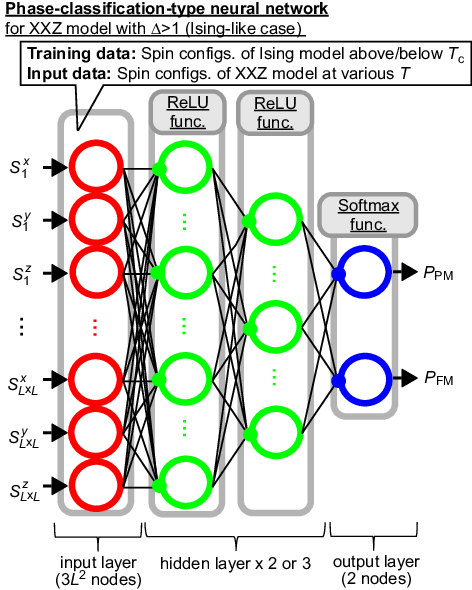}
\caption{(Color online) Basic structure of a neural network used in the phase-classification (PC) method. The output is a vector whose dimension is the number of possible phases. Each component of the vector represents the probability that the input spin or vortex configuration belongs to the corresponding phase. A lattice system of $L \times L$ sites is considered. This figure is taken and modified from Ref.~\cite{Miyajima23} {\copyright} 2023 American Physical Society.}
\label{Fig04}
\end{figure}
%%%%%%%%%%%%%%%%%%%%%%%%%%%%%%%%%%%%%%%
In the phase-classification (PC) method, the detection of phase transitions in a model to be studied is performed by solving a classification problem for other models~\cite{Kim18,Bachtis20b,Fukushima21}. Figure~\ref{Fig04} shows an example of the basic structure of a neural network used in the PC method, which is composed of one input layer, two or three hidden layers, and one output layer. The Rectified Linear Unit (ReLU) function is used as the activation function in the hidden layers, while the softmax function in the output layer. 

The ReLU function is defined by,
%%%%%%%%%%%%%%%%%%%%%%%%%%%%%%%%%%%%%%%%%%%%%%%%%%
\begin{align}
{\rm ReLU}(x)= 
\begin{cases}
x & (x \geq 0)\\ 
0 & (x < 0).
\end{cases}
\end{align}
%%%%%%%%%%%%%%%%%%%%%%%%%%%%%%%%%%%%%%%%%%%%%%%%%%
On the other hand, the softmax function is defined by,
%%%%%%%%%%%%%%%%%%%%%%%%%%%%%%%%%%%%%%%%%%%%%%%%%%
\begin{align}
{\rm Softmax}(x_i, \{x_k\}_{k=1}^N)=e^{x_i}/\sum_{k=1}^N e^{x_k},
\end{align}
%%%%%%%%%%%%%%%%%%%%%%%%%%%%%%%%%%%%%%%%%%%%%%%%%%
where $i$ and $N$ are the index of nodes and the total number of nodes in the layer. According to this expression, ${\rm Softmax}(x_i, \{x_k\}_{k=1}^N)$ is always positive, and the sum $\sum_{i=1}^N{\rm Softmax}(x_i, \{x_k\}_{k=1}^N)$ is always unity. Therefore, the output ${\rm Softmax}(x_i, \{x_k\}_{k=1}^N)$ can be regarded as probability. Because of this property, the softmax function is often used in the output layer of neural networks for classification problems. The number of nodes in each layer is set such that it is decreased equally from the input to the output layers. The input of the neural network in the PC method is spin or vortex configuration. The output is a vector whose dimension is the number of phases that may appear. For example, when we study a model which is known to undergo a single second-order (BKT) phase transition from the paramagnetic phase to the ferromagnetic (BKT) phase, the output is a two-component vector where the first component corresponds to the paramagnetic phase and the second component to the ferromagnetic (BKT) phase. The value of each component of the vector represents the probability that the input spin or vortex configuration belongs to the corresponding phase. Because the softmax function is used as the activation function in the output layer, the sum of the components of the output vector is unity. Hence the components can be interpreted as the probabilities.

For the training data, spin configurations of the Ising model can be exploited for example, when we attempt to detect the second-order phase transition from paramagnetic to ferromagnetic phases. On the other hand, when we attempt to detect the BKT transition, spin or vortex configurations of the XY model can be exploited for example. These spin and vortex configurations are generated by using the Monte Carlo thermalization technique. When two phases may appear as these examples, the answer for output is set as $(1,0)$ if the spin or vortex configuration belongs to the paramagnetic phase, while as $(0,1)$ if it belongs to the ferromagnetic or BKT phase. The neural network is trained so as to correctly classify the input spin or vortex configuration to the phase to which it belongs. 

After training, the spin or vortex configurations of the model to be studied are fed to the neural network. For input data generated at various temperatures $T$, a vector $\left(P_1(T), P_2(T)\right)$ is obtained as output when two phases are possible. We perform the same work for all temperature points and plot the obtained temperature profiles of $P_1(T)$ and $P_2(T)$. Then we can interpret the temperature at which $P_1(T)=P_2(T)=0.5$ corresponding to their intersection as the transition temperature. This is based on the consideration that the difference between the two phases disappears and the neural network cannot determine the phase at the transition temperature. To determine the transition temperature accurately, it is necessary to reduce the uncertainty of the output values $P_1(T)$ and $P_2(T)$ due to statistical fluctuations. Thus, multiple input data (typically about 100 data) are used for each temperature point, and the transition temperature is determined from averages of the output values $\overline{P}_1(T)$ and $\overline{P}_2(T)$.

Finally, we briefly explain how to prepare the input data. The input spin-configurations are represented by vectors whose components are spin components at all sites. For example, the spin configurations for three-dimensional classical spins on $L \times L$ lattice sites are given by $3L^2$-component vectors $\left(S_1^x, S_1^y, S_1^z, \cdots , S_{L \times L}^x, S_{L \times L}^y, S_{L \times L}^z \right)$. On the contrary, for the Ising model with binary spins, the spins have only one component so that they need to be extended to fit the dimension of the data. Specifically, the Ising spins are converted to three-component spins as $S_i^z \rightarrow \left(0, 0, S_i^z \right)$. The spin configurations for the Ising model are given by vectors $\left(0, 0, S_1^z, \cdots , 0, 0, S_{L \times L}^z \right)$. On the other hand, the input vortex configurations for a system of $L \times L$ sites are originally given by $L \times L$ matrices, each component of which represents the vorticity at the corresponding plaquette. To capture the spatial correlation between vortices and antivortices, a convolutional neural network is employed. Specifically, we perform a convolution operation to the vortex-configuration matrix and its vectorization before we input it to the neural network. 

A unique feature of the PC method is that it does not use the spin and vortex configurations of the target model for which the phase transition is to be detected in training the neural network. Instead, only spin and vortex configurations of models whose properties of phase transitions are well known, such as the Ising model and the XY model, are used for training. This method enables us to investigate phase transitions of unknown models using a well-known model as a steppingstone.

\subsection{Temperature-Identification (TI) method}
%%%%%%%%%%%%% Fig5 %%%%%%%%%%%%%%%%%%%%%%%
\begin{figure}[tb]
\includegraphics[scale=1.0]{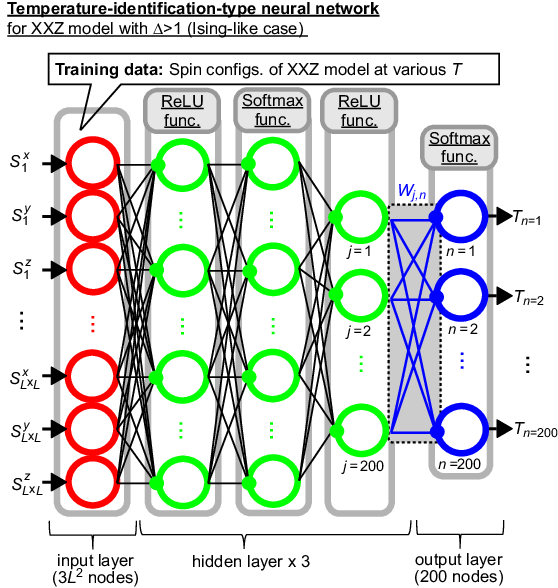}
\caption{(Color online) Basic structure of a neural network used in the temperature-identification (TI) method. The output is a vector whose dimension is the number of temperature points. Each component of the vector represents the probability that the input spin or vortex configuration is generated at the corresponding temperature range. A lattice system of $L \times L$ sites is considered. This figure is taken and modified from Ref.~\cite{Miyajima23} {\copyright} 2023 American Physical Society.}
\label{Fig05}
\end{figure}
%%%%%%%%%%%%%%%%%%%%%%%%%%%%%%%%%%%%%%%
The temperature-identification (TI) method detects phase transitions of the spin model through executing a temperature estimation task by a neural network~\cite{Tanaka17,Arai18}. Figure \ref{Fig05} shows an example of the basic structure of a neural network used in the TI method. Here, we consider a neural network consisting of one input layer, three hidden layers, and one output layer. The ReLU function is used as the activation function in the first and third hidden layers, and the softmax function in the second hidden layer and output layer. The inputs of the neural network are spin or vortex configurations. The output is an $N$-component vector ($N$=200-300), where its $n$th component corresponds to the temperature point $T_n=n\Delta T$ $(n=1, 2, \cdots, N)$. Specifically, its value is regarded as the probability that the input spin or vortex configuration is generated at temperature $T_n$.

The training data are spin or vortex configurations generated by the Monte Carlo thermalization technique. The answers for outputs are vectors representing the temperature at which the input spin or vortex configuration was generated. Specifically, if the spin or vortex configuration generated at temperature $T_{n}$ is input, only the $n$th component of the vector is set to be unity while the other components to be zero. This representation is called one-hot representation. The input data are the same as those for the PC method. The neural network is trained so as to correctly estimate the temperature at which the input spin or vortex configuration is generated.

After the training, we focus on the weight matrix $\hat{W}$ connecting the last hidden layer and the output layer. Its component $W_{j,n}$ represents the weight connecting the $j$th node of the last hidden layer and the $n$th node of the output layer, where the $n$th column corresponds to the temperature $T_n=n \Delta T$. This weight matrix contains information about phase transitions of the model, and thus the transition temperatures can be determined by analyzing the heat map. Here the heat map is a grayscale plot of the weight-matrix components by regarding them as pixel values on a plane with horizontal axis $n$ $(=1, 2, \cdots, N)$ and vertical axis $j$ $(=1, 2, \cdots, N_{\rm h})$ where $N_{\rm h}$ is the number of nodes in the last hidden layer. It is known that the heat map often changes its pattern at $n$ corresponding to the transition temperature $T_n$.

To analyze the pattern change of the heat map quantitatively, the correlation function $C_W(T)$ and variance $V_W(T)$ were proposed~\cite{Miyajima21,Miyajima23}. The correlation function $C_W(T)$ is defined by,
%%%%%%%%%%%%%%%%%%%%%%%%%%%%%%%%%%%%%%%%%%%%%%%%%%
\begin{align}
C_W(T_n) = \frac{1}{N_{\rm h}} \sum_{m=1}^{n-1} \sum_{j=1}^{N_{\rm h}} W_{j,m}W_{j,m+1}.
\end{align}
%%%%%%%%%%%%%%%%%%%%%%%%%%%%%%%%%%%%%%%%%%%%%%%%%%
This formula implies that the inner products taken between adjacent columns in the transversal direction of the weight matrix are cumulatively summed up from the low-temperature side. The product $W_{j,m}W_{j,m+1}$ is a quantity that quantifies how close adjacent columns in the heat map are. When the pattern of the heat map changes at $T_n$, the slope of $C_W(T) $ is expected to change at this temperature. On the other hand, the variance $V_W(T)$ is defined by,
%%%%%%%%%%%%%%%%%%%%%%%%%%%%%%%%%%%%%%%%%%%%%%%%%%
\begin{align}
V_W(T_n) =  \frac{1}{n-1} \sum_{j=1}^{N_h} \sum_{m=1}^{n-1} (W_{j,m} - \overline{W_j})^2,
\end{align}
%%%%%%%%%%%%%%%%%%%%%%%%%%%%%%%%%%%%%%%%%%%%%%%%%%
with
%%%%%%%%%%%%%%%%%%%%%%%%%%%%%%%%%%%%%%%%%%%%%%%%%%
\begin{align}
\overline{W_j} = \frac{1}{n-1} \sum_{m=1}^{n-1} W_{j,m}.
\end{align}
%%%%%%%%%%%%%%%%%%%%%%%%%%%%%%%%%%%%%%%%%%%%%%%%%%
This formula implies that we take the variance of the components corresponding to the temperature points $T=T_1, T_2, \cdots, T_{n-1}$ in each row of the weight matrix and sum them up over all rows. If no phase transition occurs in the temperature range $[T_1, T_{n-1}]$, the value of $V_W(T)$ is expected to be small because the components for taking the variance $V_W(T)$ are close to each other. On the other hand, if a phase transition occurs in the temperature range $[T_1, T_{n-1}]$, the value of $V_W(T_n)$ is expected to increase rapidly because some components for taking the variance $V_W(T)$ have significantly different values. Therefore, the transition temperature can be determined from the point where the variance $V_W(T)$ plotted against the temperature $T$ begins to increase.

A unique feature of the TI method is that it does not require prior knowledge of phase transitions, e.g., transition temperatures and the number of possible phases, of the target model for training. The answers of output for training data are vectors representing the temperature at which the input spin or vortex configuration is generated. Namely, the answers are computational conditions of the Monte Carlo thermalization calculations and thus are known. Another unique feature is that the neural network of the TI method is trained only and no estimation by data input is performed after training.

\section{$q$-state clock models}
To demonstrate the TI method introduced in the previous section, we discuss the detection of a single second-order phase transition and successive BKT transitions in the $q$-state clock models.

\subsection{Single second-order phase transition when $q$=4}
%%%%%%%%%%%%% Fig6 %%%%%%%%%%%%%%%%%%%%%%%
\begin{figure}[tb]
\includegraphics[scale=0.5]{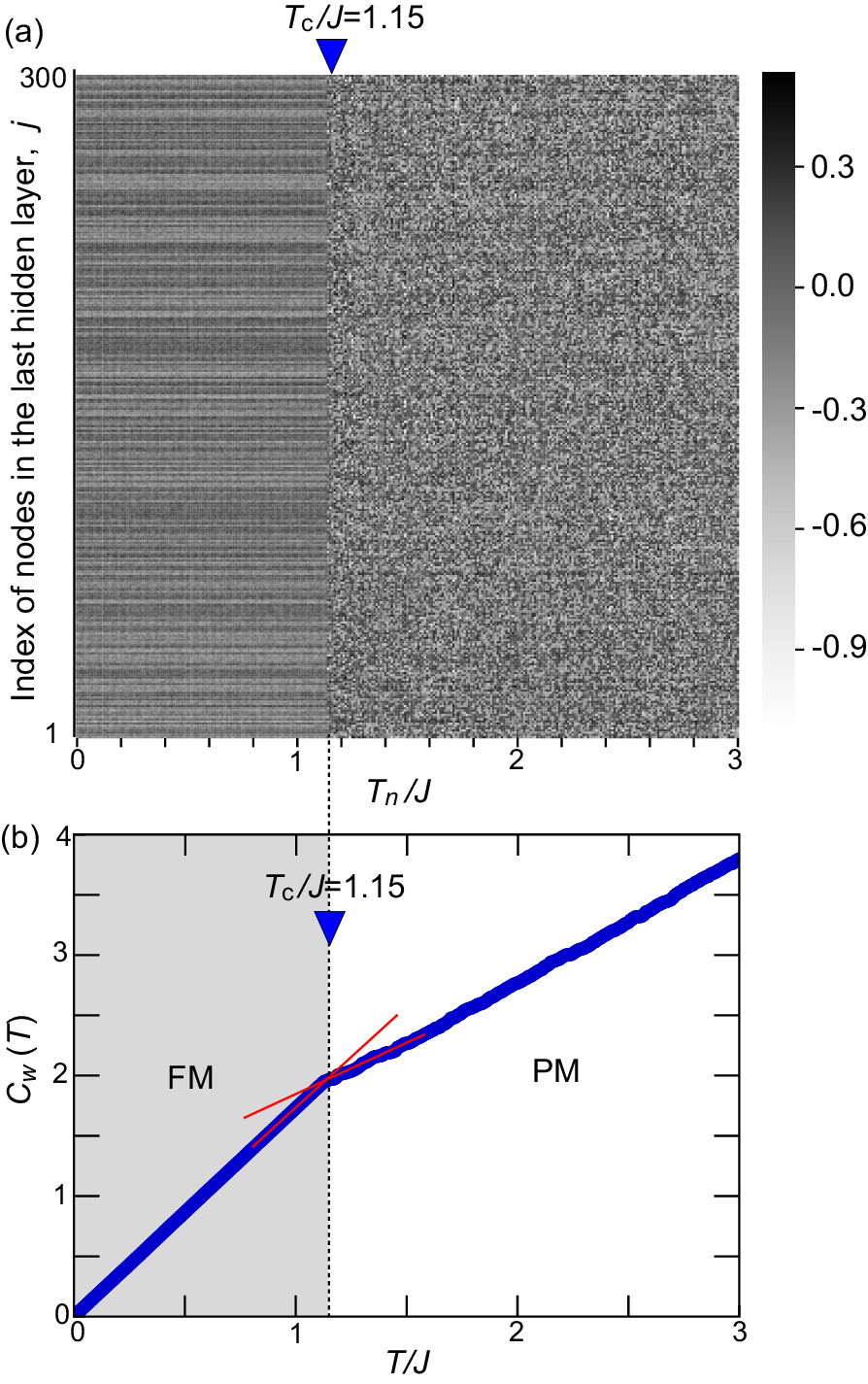}
\caption{(Color online) (a) Heat map of the weight matrix connecting the last hidden layer and the output layer of the neural network in the TI method for the four-state ($q=4$) clock model. The neural network is trained using the spin-configuration data of the four-state clock model for $L$=80. A pattern change from the horizontal-stripe pattern to the sandstorm pattern can be seen, which indicates the single second-order phase transition. (b) Temperature profile of the correlation function $C_W(T)$. From the change of its slope, the transition temperature is evaluated to be $T_{\rm c}/J$=1.15. This figure is taken and modified from Ref.~\cite{Miyajima21} {\copyright} 2021 American Physical Society.}
\label{Fig06}
\end{figure}
%%%%%%%%%%%%%%%%%%%%%%%%%%%%%%%%%%%%%%%
We first discuss the detection of the single second-order phase transition in the four-state ($q=4$) clock model using the TI method. First, a large number of spin configurations are generated by the Monte Carlo thermalization technique. Specifically, 10 spin configurations for a lattice size $L=80$ at each of 300 temperature points $T_n=n \Delta T$ $(n=1, 2, \cdots, 300)$ are prepared. Second, the neural network is trained using these spin configurations. The weight matrices connecting adjacent layers are optimized so as to correctly estimate the temperature at which the spin configuration was generated. Third, the weight matrix connecting the last hidden layer and the output layer of the optimized neural network is analyzed to determine the transition temperature. Figure~\ref{Fig06}(a) shows a heat map of the weight matrix. The horizontal axis represents the index of nodes $n$ in the output layer corresponding the temperature $T_n/J$, while the vertical axis represents the index of nodes $j$ in the last hidden layer. A change of pattern is clearly discernible in this heat map. A horizontal-stripe pattern appears in the low-temperature regime, while a sandstorm pattern appears in the high-temperature regime. From this pattern change, the transition temperature for the four-state clock model can be evaluated as $T_{\rm c}/J$=1.15. This value is in good agreement with the exact value of $T_{\rm c}/J=1/\ln(1+\sqrt{2})\approx 1.1346$ in the thermodynamic limit~\cite{Onsager44,Brush67}. Because the pattern change is clear in this case, we can achieve the accurate evaluation by visual inspection without calculating the correlation function $C_W(T)$. The temperature profile of $C_W(T)$ is shown in Fig.~\ref{Fig06}(b), which shows a clear change in slope at the transition temperature $T_{\rm c}/J=1.15$, indicating that this quantity captures the pattern change in the heat map with high accuracy.

\subsection{Successive BKT transitions when $q$=8}
%%%%%%%%%%%%% Fig7 %%%%%%%%%%%%%%%%%%%%%%%
\begin{figure}[tb]
\includegraphics[scale=0.5]{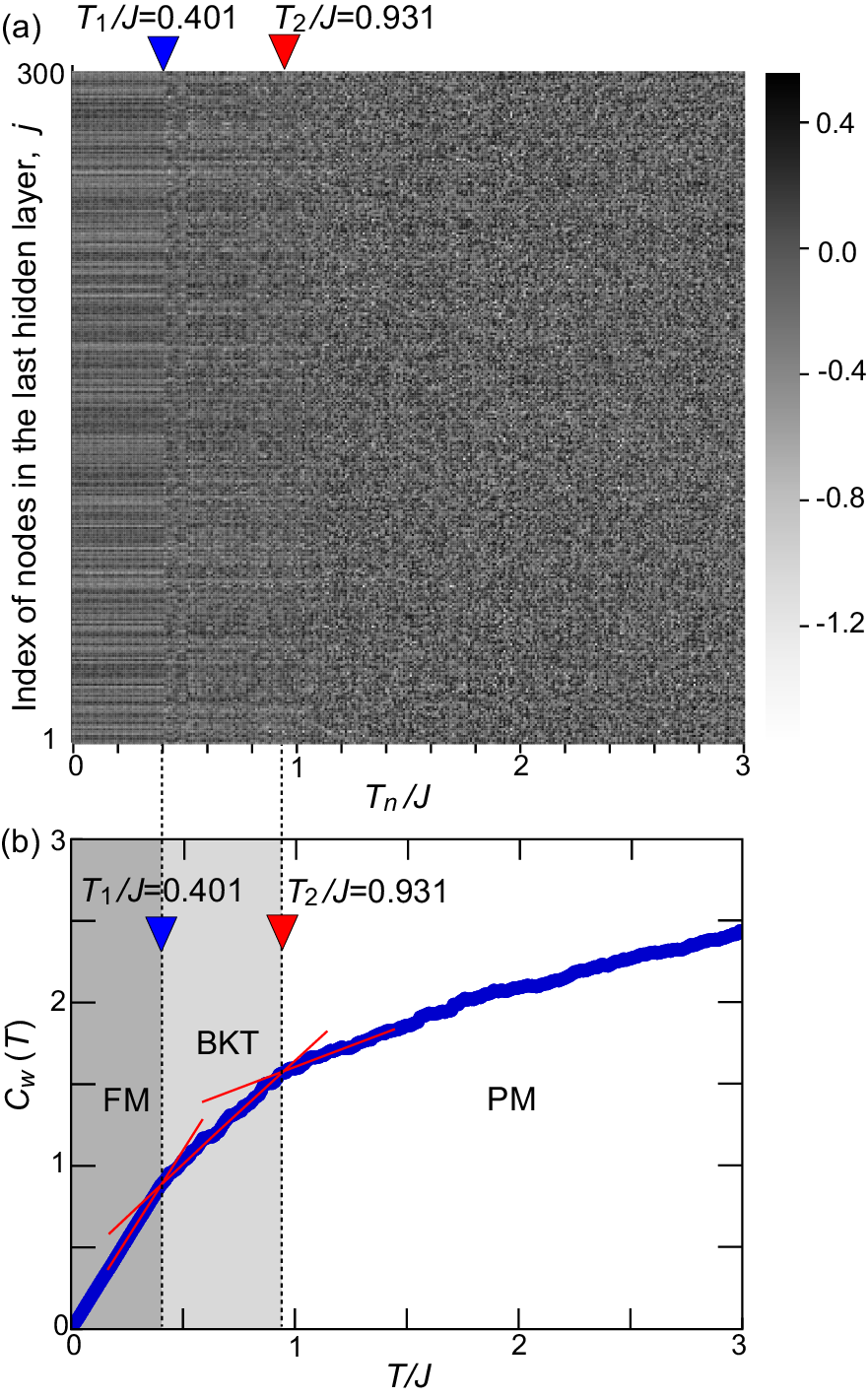}
\caption{(Color online) (a) Heat map of the weight matrix connecting the last hidden layer and the output layer of the neural network in the TI method for the eight-state ($q=8$) clock model. The neural network is trained using the spin-configuration data of the eight-state clock model for $L$=72. (b) Temperature profile of the correlation function $C_W(T)$. The slope changes twice corresponding to the successive two BKT transitions. This figure is taken and modified from Ref.~\cite{Miyajima21} {\copyright} 2021 American Physical Society.}
\label{Fig07}
\end{figure}
%%%%%%%%%%%%%%%%%%%%%%%%%%%%%%%%%%%%%%%
%%%%%%%%%%%%%%%%%%%%%%%%%%%%%%%%%%%%%%%%%%%%%%%%%%%
\begin{table}[tbh]%[h]
\caption{BKT-transition temperatures of the eight-state ($q=8$) clock model obtained for various lattice sizes. Here $L \rightarrow \infty$ denotes the thermodynamic limit.}
\begin{tabular}{cccccccc}
\hline
$L$ & 16 & 32 & 48 & 64 & 72 & 80 & $\infty$ \\
\hline 
$T_1/J$ & 0.319 & 0.376 & 0.388 & 0.392 & 0.401 & 0.390 & 0.410 \\
$T_2/J$ & 1.279 & 1.000 & 0.943 & 0.931 & 0.931 & 0.930 & 0.921 \\
\hline
\end{tabular}
\label{Table01}
\end{table}
%%%%%%%%%%%%%%%%%%%%%%%%%%%%%%%%%%%%%%%%%%%%%%%%%%%
We next discuss the detection of the two BKT transitions in the eight-state ($q=8$) clock model using the TI method. Again, 10 spin configurations for $L$=72 are generated at each of 300 temperature points $T_n=n \Delta T$ $(n=1, 2, \cdots, 300)$ by the Monte Carlo thermalization technique for training data. Figure~\ref{Fig07}(a) shows a heat map of the weight matrix connecting the last hidden layer and the output layer of the trained neural network. In this figure, a pattern change associated with the BKT transition at lower temperature, which corresponds to the first BKT transition between the BKT phase and the ferromagnetic phase, can be seen around $T_n/J=0.4$. Another pattern change can also be seen at higher temperature, which is associated with the second BKT transition between the paramagnetic phase and the BKT phase. However, it is hard to correctly determine the boundary of the pattern change by human eyes. Now, the correlation function $C_W(T)$ enables us quantitative clarification of the pattern changes. Figure~\ref{Fig07}(b) shows the temperature profile of $C_W(T)$. In this figure, we see that there are three phases characterized by different values of slope of this profile. Namely, the slope of $C_W(T)$ changes twice, which correspond to the successive two BKT transitions. Their transition temperatures can be evaluated at the same time from the profile of $C_W(T)$. By performing the same analyses for various lattice sizes, the transition temperatures in the thermodynamic limit are obtained by extrapolation in the finite-size scaling analysis. The results are summarized in Table~\ref{Table01}.

%%%%%%%%%%%%%%%%%%%%%%%%%%%%%%%%%%%%%%%%%%%%%%%%%%%
\begin{table}[tbh]%[h]
\caption{Comparison of the BKT-transition temperatures in the thermodynamic limit obtained by the TI method in Ref.~\cite{Miyajima21} and Monte Carlo methods in previous studies~\cite{Tomita02,Brito10}.}
\begin{tabular}{lll}
\hline
Reference & $T_1 / J$ & $T_2 / J$ \\ \hline 
Ref.~\cite{Miyajima21} (TI method) & 0.410 & 0.921 \\ 
Ref.~\cite{Tomita02}   & 0.4259(4) & 0.8936(7) \\
Ref.~\cite{Brito10}      & 0.42(2)     & 0.898(7)  \\
\hline
\end{tabular}
\label{Table02}
\end{table}
%%%%%%%%%%%%%%%%%%%%%%%%%%%%%%%%%%%%%%%%%%%%%%%%%%%
%%%%%%%%%%%%% Fig8 %%%%%%%%%%%%%%%%%%%%%%%
\begin{figure}[tb]
\includegraphics[scale=1.0]{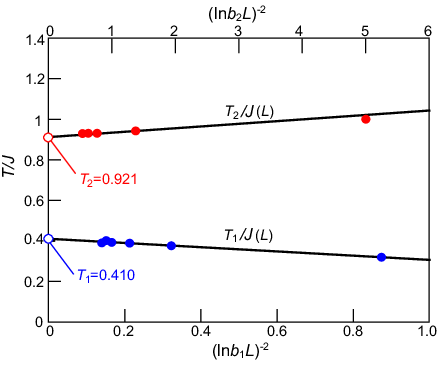}
\caption{(Color online) System-size scaling of the two transition temperatures $T_1$ and $T_2$ for the eight-state ($q$=8) clock model. The values in the thermodynamic limit are obtained as $T_1/J$=0.410 and $T_2/J$=0.921 by extrapolations (see text). This figure is taken and modified from Ref.~\cite{Miyajima21} {\copyright} 2021 American Physical Society.}
\label{Fig08}
\end{figure}
%%%%%%%%%%%%%%%%%%%%%%%%%%%%%%%%%%%%%%%
According to the theory developed by Kosterlitz and Thouless, the correlation length $\xi$ is given by $b\exp(c/\sqrt{t})$ on the verge of the BKT transition where $b$ and $c$ are constants. The relative temperature $t$ is given using the BKT-transition temperature in the thermodynamic limit $T_{\rm BKT}(\infty)$ as,
%%%%%%%%%%%%%%%%%%%%%%%%%%%%%%%%%
\begin{align}
t=\frac{|T-T_{\rm BKT}(\infty)|}{T_{\rm BKT}(\infty)}.
\end{align}
%%%%%%%%%%%%%%%%%%%%%%%%%%%%%%%%%
Because the correlation length $\xi$ becomes identical to the system size $L$ (i.e., $\xi=L$) at $T=T_{\rm BKT}(L)$ in finite-size systems, the system-size dependence of $T_{\rm BKT}(L)$ is given by,
%%%%%%%%%%%%%%%%%%%%%%%%%%%%%%%%%
\begin{align}
T_{\rm BKT}(L) = T_{\rm BKT}(\infty) \pm \frac{c^2 T_{\rm BKT}(\infty)}{(\ln bL)^2}.
\end{align}
%%%%%%%%%%%%%%%%%%%%%%%%%%%%%%%%%
Here the sign of the second term should be $+$ (plus) for the higher transition temperature $T_2$ and $-$ (minus) for the lower transition temperature $T_1$. Extrapolations using this equation give the transition temperatures in the thermodynamic limit as $T_1/J$=0.410 and $T_2/J$=0.921 (see Fig.~\ref{Fig08}). These values are in good agreement with those obtained by the Monte Carlo methods previously (see Table~\ref{Table02}).

\section{Second-order transition in the Ising-like XXZ model}
%\subsection{Detection by the PC method}
%%%%%%%%%%%%% Fig9 %%%%%%%%%%%%%%%%%%%%%%%
\begin{figure}[tb]
\includegraphics[scale=1.0]{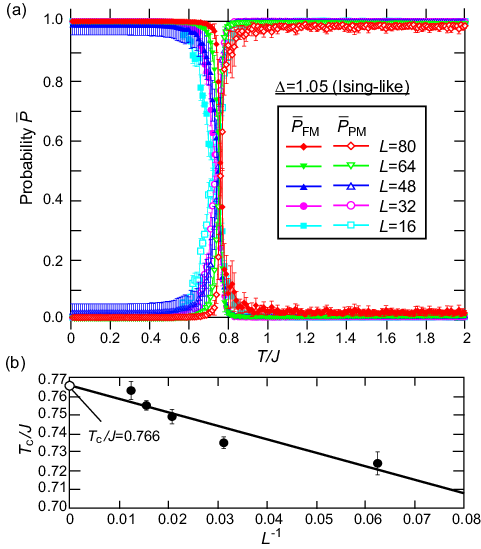}
\caption{(Color online) (a) Temperature profiles of the averaged outputs $\overline{P}_{\rm FM}$ and $\overline{P}_{\rm PM}$ of the neural network in the PC method for various lattice sizes, which are obtained by feeding spin configurations of the Ising-like XXZ model with $\Delta$=1.05 as input data to the neural network trained using spin configurations of the Ising model as training data. The spin configurations are generated by using the Monte Carlo thermalization technique. The transition temperature is evaluated from the intersection of $\overline{P}_{\rm FM}$ and $\overline{P}_{\rm PM}$ plots for each lattice size. (b) Finite-size scaling analysis of the transition temperatures $T_{\rm c}$. The transition temperature in the thermodynamic limit is evaluated as $T_{\rm c}/J$=0.766 by extrapolation, which is in good agreement with $T_{\rm c}/J$=0.76 obtained by the Monte Carlo method. This figure is taken and modified from Ref.~\cite{Miyajima23} {\copyright} 2023 American Physical Society.}
\label{Fig09}
\end{figure}
%%%%%%%%%%%%%%%%%%%%%%%%%%%%%%%%%%%%%%%
%%%%%%%%%%%%%%%%%%%%%%%%%%%%%%%%%%%%%%%%%%%%%%%%%%%
\begin{table*}[tbh]%[h]
\caption{Comparison between the transition temperatures obtained by the Monte Carlo method $T_{\rm c}^{\rm MC}$ and those obtained by the machine learning analysis with the PC method $T_{\rm c}^{\rm PC}$ for the Ising-like XXZ model with $\Delta$=1.05 for various lattice sizes. Here $L \rightarrow \infty$ denotes the thermodynamic limit.}
\begin{tabular}{ccccccc}
\hline
$L$ & 16 & 32 & 48 & 64 & 80 & $\infty$ \\
\hline 
$T_{\rm c}^{\rm MC}/J$ & 0.760(5) & 0.760(5) & 0.760(5) & 0.760(5) & 0.760(5) & 0.760(0) \\
$T_{\rm c}^{\rm PC}/J$ & 0.724(6) & 0.735(3) & 0.749(4) & 0.755(3) & 0.763(6) & 0.766(5) \\
\hline
\end{tabular}
\label{Table03}
\end{table*}
%%%%%%%%%%%%%%%%%%%%%%%%%%%%%%%%%%%%%%%%%%%%%%%%%%%
Here we discuss the detection of the second-order phase transition in the Ising-like XXZ model with $\Delta$=1.05 using the PC method. The neural network is trained using spin configurations of the Ising model. Using the Monte Carlo thermalization technique, 100 spin configurations are generated at each of the 400 temperature points $T_n=n\Delta T$ $(n=1, 2, \cdots, 400)$. To avoid explicitly learning critical properties, the spin configurations of the Ising model around the transition temperature $T_{\rm c}/J=2/\log(\sqrt{2}+1)\approx 2.269$~\cite{Onsager44} are not used for training. After training, 100 spin configurations of the XXZ model with $\Delta$=1.05 generated at each of 200 temperature points $T_n=n \Delta T$ $(n=1, 2, \cdots, 200)$ are fed to the optimized neural network. The averaged probabilities $\left(\overline{P}_{\rm PM}, \overline{P}_{\rm FM} \right)$ over the 100 outputs obtained at each temperature point are used for detection of the phase transition. 

Figure~\ref{Fig09}(a) shows the temperature profiles of $\overline{P}_{\rm PM}$ and $\overline{P}_{\rm FM}$ for various lattice sizes. At higher temperatures, $\overline{P}_{\rm FM}$ is almost zero, while $\overline{P}_{\rm PM}$ is nearly unity. On the contrary, at lower temperatures, the relationship is reversed, that is, $\overline{P}_{\rm FM}$ becomes nearly unity, while $\overline{P}_{\rm PM}$ becomes suppressed to be zero. This contrasting behaviors between two temperature ranges indicates that the neural network correctly recognizes the two phases in this model. The transition temperature is evaluated from the intersection of the two temperature-profiles. Table~\ref{Table03} shows the transition temperatures obtained by the Monte Carlo method and the PC method for various lattice sizes for comparison. The finite-size scaling analysis results in the transition temperature $T_{\rm c}/J$=0.766 in the thermodynamic limit for the PC method [Fig.~\ref{Fig09}(b)]. This value is in good agreement with $T_{\rm c}/J$=0.76 obtained by specific-heat data calculated by the Monte Carlo method.

%\subsection{Detection by the TI method}
%%%%%%%%%%%%% Fig10 %%%%%%%%%%%%%%%%%%%%%%%
\begin{figure}[tb]
\includegraphics[scale=0.5]{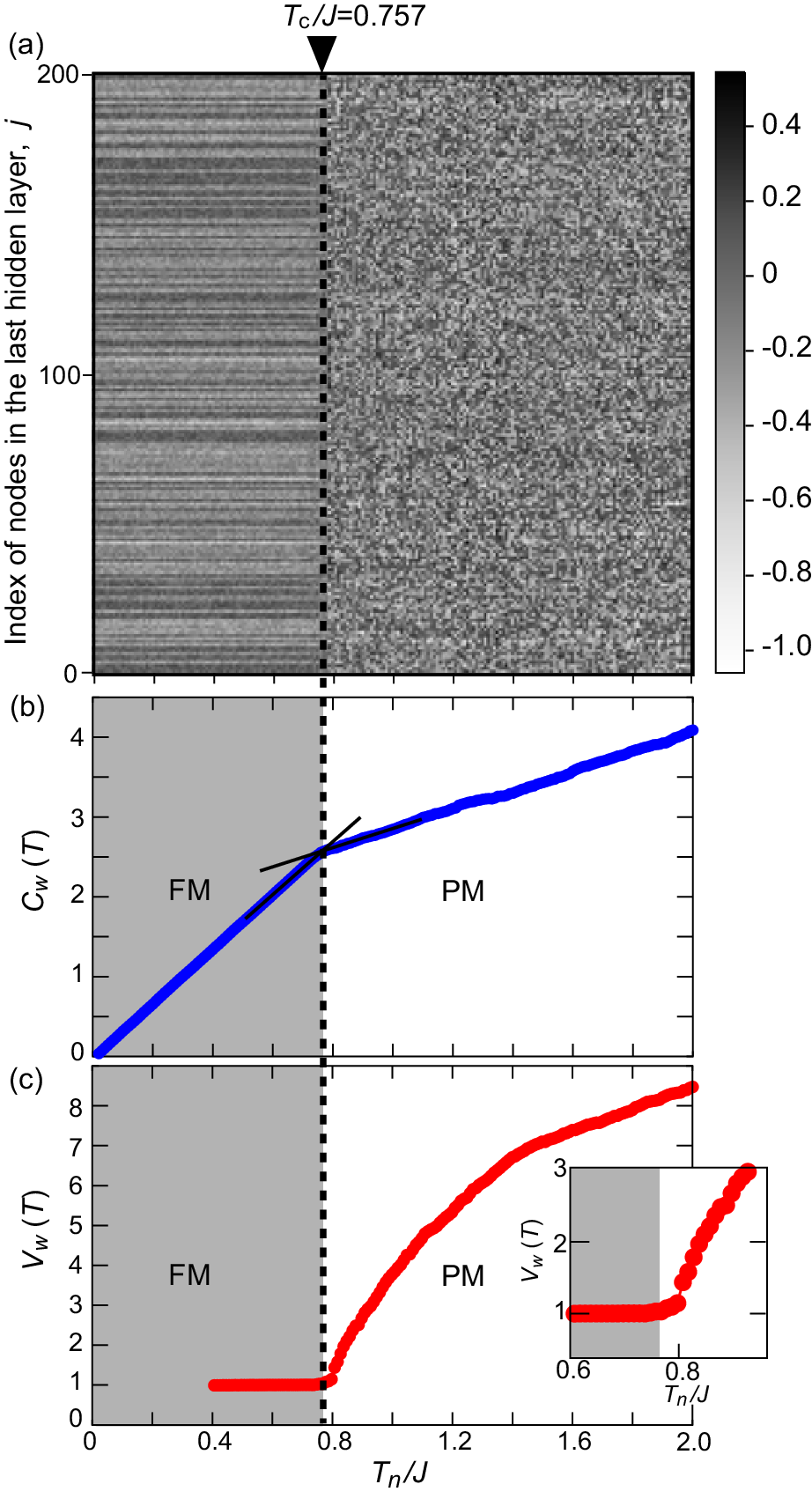}
\caption{(Color online) (a) Heat map of the weight matrix connecting the last hidden layer and the output layer of the neural network in the TI method. The neural network is trained using the spin-configuration data of the Ising-like XXZ model with $\Delta$=1.05 for $L$=72. A pattern change from the horizontal-stripe pattern to the sandstorm pattern can be seen, which indicates the second-order phase transition. (b) Temperature profile of the correlation function $C_W(T)$. From the change of its slope, the transition temperature is evaluated to be $T_{\rm c}/J$=0.757. (c) Temperature profile of the variance $V_W(T)$. From its abrupt rise, the transition temperature is evaluated to be $T_{\rm c}/J$=0.75. Two evaluated transition temperatures are in good coincidence. This figure is taken and modified from Ref.~\cite{Miyajima23} {\copyright} 2023 American Physical Society.}
\label{Fig10}
\end{figure}
%%%%%%%%%%%%%%%%%%%%%%%%%%%%%%%%%%%%%%%
We then discuss the detection of the second-order phase transition in the Ising-like XXZ model with $\Delta$=1.05 using the TI method. For training the neural network, 100 spin configurations of this model for $L$=80 are generated at each of 200 temperature points $T_n=n \Delta T$ $(n=1, 2, \cdots, 200)$ by the Monte Carlo thermalization technique. Figure~\ref{Fig10}(a) shows a heat map of the weight matrix connecting the last hidden layer and the output layer of the neural network after training. In this figure, a horizontal-stripe pattern appears at lower temperatures, while a sandstorm pattern appears at higher temperatures. The pattern change occurs at around $T_n/J$=0.76. The correlation function $C_W(T)$ and variance $V_W(T)$ are calculated from this weight matrix. The temperature profile of $C_W(T)$ is shown in Fig.~\ref{Fig10}(b), which exhibits a change of slope corresponding to the second-order phase transition at $T_{\rm c}^{\rm cf}/J$=0.757. On the other hand, the temperature profile of $V_W(T)$ is shown in Fig.~\ref{Fig10}(c), which exhibits an abrupt increase again corresponding to the second-order phase transition at $T_{\rm c}^{\rm var}/J$=0.75. Both the transition temperatures evaluated from $C_W(T)$ and $V_W(T)$ are in good agreement with the transition temperature obtained by the Monte Carlo method. Note that the $q$-state clock models discussed in the previous section have discretized two-dimensional spins, while the XXZ models have continuous three-dimensional spins. It seems that this difference does not affect the detection of the second-order phase transition.

\section{BKT transition in the XY-like XXZ model}
%\subsection{Detection by the PC method}
%%%%%%%%%%%%%% Fig11 %%%%%%%%%%%%%%%%%%%%%%%
\begin{figure}[tb]
\includegraphics[scale=1.0]{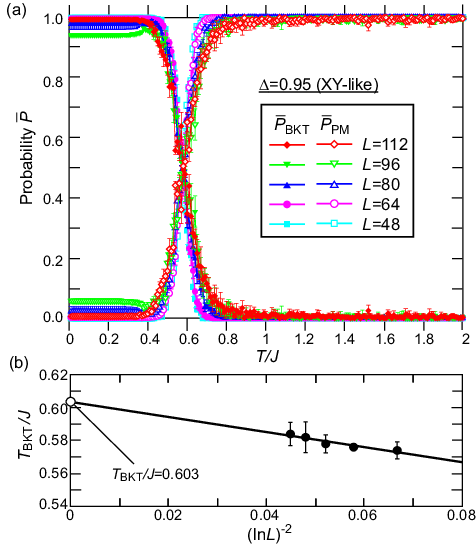}
\caption{(Color online) (a) Temperature profiles of the averaged outputs $\overline{P}_{\rm PM}$ and $\overline{P}_{\rm BKT}$ of the neural network in the PC method for various lattice sizes, which are obtained by feeding vortex configurations of the XY-like XXZ model with $\Delta$=0.95 as input data to the neural network trained using vortex configurations of the XY model. The vortex configurations are generated by using the Monte Carlo thermalization technique. The transition temperature is evaluated from the intersection of $\overline{P}_{\rm PM}$ and $\overline{P}_{\rm BKT}$ plots for each lattice size. (b) Finite-size scaling analysis of the transition temperatures $T_{\rm BKT}$. The transition temperature in the thermodynamic limit is evaluated as $T_{\rm BKT}/J$=0.603 by extrapolation, which is in good agreement with $T_{\rm BKT}/J$=0.606 obtained by the Monte Carlo method. This figure is taken and modified from Ref.~\cite{Miyajima23} {\copyright} 2023 American Physical Society.}
\label{Fig11}
\end{figure}
%%%%%%%%%%%%%%%%%%%%%%%%%%%%%%%%%%%%%%%
%%%%%%%%%%%%%%%%%%%%%%%%%%%%%%%%%%%%%%%%%%%%%%%%%%%
\begin{table*}[tbh]%[h]
\caption{Comparison between the BKT-transition temperatures obtained by the PC method $T_{\rm BKT}^{\rm PC}$ and those obtained by the Monte Carlo method $T_{\rm BKT}^{\rm MC}$ for the XY-like XXZ model with $\Delta$=0.95 for various lattice sizes. Here $L \rightarrow \infty$ denotes the thermodynamic limit.}
\begin{tabular}{ccccccccc} 
\hline
$L$ & 16 & 32 & 48 & 64 & 80 & 96 & 112 & $\infty$ \\
\hline
$T_{\rm BKT}^{\rm PC}/J$ & - & - & 0.574(5) & 0.576(2) & 0.578(5) & 0.582(9) & 0.584(7) & 0.603(5) 
\\
\\
$T_{\rm BKT}^{\rm MC}/J$ & 0.6351 & 0.6243 & 0.6201 & 0.6188 & 0.6177 & - & - & 0.606 \\
\hline
\end{tabular}
\label{Table04}
\end{table*}
%%%%%%%%%%%%%%%%%%%%%%%%%%%%%%%%%%%%%%%%%%%%%%%%%%%
We discuss the detection of the BKT transition in the XY-like XXZ model with $\Delta$=0.95 using the PC method. The neural network is trained using vortex configurations of the XY model. Using the Monte Carlo thermalization technique, 100 vortex configurations are generated at each of the 200 temperature points $T_n=n\Delta T$ $(n=1, 2, \cdots, 200)$. After training, 100 vortex configurations of the XXZ model with $\Delta$=0.95 generated at each of 200 temperature points $T_n=n \Delta T$ $(n=1, 2, \cdots, 200)$ are fed to the optimized convolutional neural network. The averaged probabilities $\left(\overline{P}_{\rm PM}, \overline{P}_{\rm BKT} \right)$ over the 100 outputs obtained at each temperature point are used for detection of the phase transition. 

Figure~\ref{Fig11}(a) shows the temperature profiles of $\overline{P}_{\rm PM}$ and $\overline{P}_{\rm BKT}$ for various lattice sizes. At low temperatures $\overline{P}_{\rm PM}\approx 0$ and $\overline{P}_{\rm BKT}\approx 1$, while at high temperatures $\overline{P}_{\rm PM}\approx 1$, $\overline{P}_{\rm BKT}\approx 0$. These contrasting behaviors between two temperature ranges indicates that the neural network correctly recognizes the two phases in this model. The transition temperature is again evaluated from the intersection of the two temperature profiles. Table~\ref{Table04} shows the transition temperatures obtained by the Monte Carlo method and the PC method for various lattice sizes for comparison. The finite-size scaling analysis results in the transition temperature $T_{\rm BKT}/J$=0.603 in the thermodynamic limit for the PC method [Fig.~\ref{Fig11}(b)]. This value is in good agreement with $T_{\rm BKT}/J$=0.606 obtained from the helicity modulus data calculated by the Monte Carlo method.

%\subsection{Detection by the TC method}
%%%%%%%%%%%%% Fig12 %%%%%%%%%%%%%%%%%%%%%%%
\begin{figure}[tb]
\includegraphics[scale=1.0]{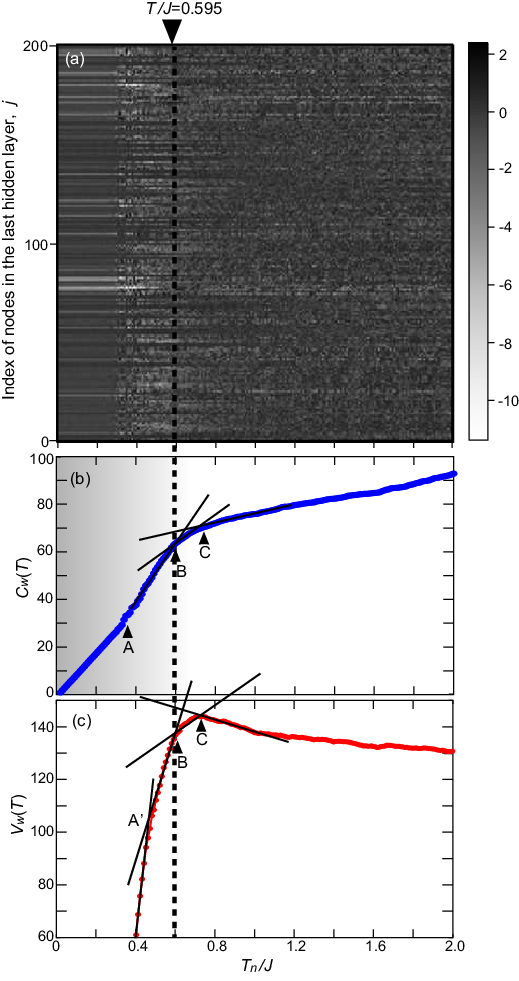}
\caption{(Color online) (a) Heat map of the weight matrix connecting the last hidden layer and the output layer of the convolutional neural network in the TI method. The neural network is trained using the vortex-configuration data of the XY-like XXZ model with $\Delta$=0.95 for $L$=80. (b) Temperature profile of the correlation function $C_W(T)$. The slope changes three times at points A, B, and C. (c) Temperature profile of the variance $V_W(T)$. The behavior changes at points A$^\prime$, B, and C. This figure is taken and modified from Ref.~\cite{Miyajima23} {\copyright} 2023 American Physical Society.}
\label{Fig12}
\end{figure}
%%%%%%%%%%%%%%%%%%%%%%%%%%%%%%%%%%%%%%%
Finally, we discuss the detection of the BKT transition in the XY-like XXZ model with $\Delta$=0.95 using the TI method. For training data, 100 vortex configurations of this model for $L$=80 are generated at each of 200 temperature points $T_n=n \Delta T$ $(n=1, 2, \cdots, 200)$ by the Monte Carlo thermalization technique. Figure~\ref{Fig12}(a) shows a heat map of the weight matrix connecting the last hidden layer and the output layer of the convolutional neural network after training. It seems that a horizontal-stripe pattern and a sandstorm pattern appear again at lower and higher temperatures, respectively. However, it is difficult to recognize their boundary by eyes because the two patterns are somehow merged at intermediate temperatures. 

To quantify the pattern changes, the correlation function $C_W(T)$ and the variance $V_W(T)$ are again calculated. The temperature profiles of $C_W(T)$ and $V_W(T)$ are shown in Figs.\ref{Fig12}(b) and (c), respectively. The profile of $C_W(T)$ shows changes in slope at three points A, B, and C. On the other hand, the profile $V_W(T)$ shows changes in slope at three points A$^\prime$, B, and C. Here the point A for $C_W(T)$ and the point A$^\prime$ for $V_W(T)$ appear at similar temperatures. The numbers of these slope changes are greater than the number of phase transitions. The heat map exhibits pattern changes even at points which are not directly related with the BKT transition, although a pattern change might appear at the real transition point. This makes it difficult to determine the BKT-transition temperature uniquely.

According to the temperature profile of vortex density calculated using the Monte Carlo method (not shown)~\cite{Miyajima23}, it seemingly begins to rise from zero at a temperature around the points A and A$^\prime$. On the other hand, the point C corresponds to an inflection point of the vortex-density profile. Certainly, these points are not associated with the BKT transition directly. The convolutional neural network seems to detect the emergence of vortices and antivortices at point A, while a kind of crossover phenomenon at point C. It is expected that a phase-transition-like behavior characterized by anomalies of certain thermodynamic quantities such as a peak of the specific heat should appear at point C. On the contrary, point B is apparently related with the BKT transition. Note that as discussed in the previous section, the BKT transitions in the eight-state clock model with discretized planar spins can be detected by the TI method using the spin configurations as input data. On the contrary, the BKT transition in the XY-like XXZ model with continuous three-dimensional spins is relatively difficult to detect by the TI method even though the vortex configurations are used as input data.

\section{Discussion}
\subsection{Comparison with other machine learning methods}
%%%%%%%%%%%%%%%%%%%%%%%%%%%%%%%%
\begin{table*}[tbh]%[h]
\caption{
Summary of the proposed machine learning methods for detection of the BKT transition.\\
*1 Spatial vortex configurations are used as the input data. \\
*2 Spin-orientation histograms are used as the input data. \\
*3 Spin correlation functions are used as the input data.}
%$*1$ Prior knowledge about the transition temperatures and the number of phases is required. $*2$ Prior knowledge about the number of phases is required.
\begin{tabular}{lllll} 
\hline
Ref. & Model & Type & Prior knowledge &  Feature engineering \\ 
\hline 
Ref.~\cite{Richter-Laskowska18}   & XY    & Supervised   & Necessary (Transition temperatures, Number of phases) & Not necessary \\
Ref.~\cite{Beach18}     & XY    & Supervised   & Necessary (Transition temperatures, Number of phases)    & Necessary$^{*1}$\\
Ref.~\cite{ZhangW19}     & XY    & Supervised   & Necessary (Number of phases) & Necessary$^{*2}$\\
Ref.~\cite{Rodriguez-Nieva19} & XY    & Unsupervised (Diffusion map method) & Not Necessary &  Not necessary\\
Ref.~\cite{Shiina20}    & Clock & Supervised   & Necessary (Transition temperatures, Number of phases)    & Necessary$^{*3}$\\
\hline
Ref.~\cite{Mendes-Santos21}& -- & Unupervised  (Intrinsic dimension method) & Not Necessary  & Not necessary\\
Ref.~\cite{Miyajima21} & Clock & Supervised  (TI method) & Not Necessary  & Not necessary\\
Ref.~\cite{Miyajima23} & XXZ  & Supervised  (TI method) & Not Necessary  & Not necessary\\
Ref.~\cite{Miyajima23} & XXZ  & Supervised   (PCmethod) & Not Necessary  & Not necessary\\
\hline
\end{tabular}
\label{Table05}
\end{table*}
%%%%%%%%%%%%%%%%%%%%%%%%%%%%%%%%
For the machine-learning detection of BKT transitions, numerous methods and their demonstrations have been reported so far. In Table~\ref{Table05}, previous methods and their features are summarized. According to this table, we find that most of the previous methods are based on supervised learning. They require either prior knowledge of the model or feature engineering of the data or even both. Some methods need to know approximate values of transition temperatures of the model in advance to prepare the training data. Namely, they use spin or vortex configurations generated at several temperatures for training, each of which is labeled by a name of the corresponding phase according to the knowledge of the transition temperatures. This means that although our aim is to detect unknown phase transitions in a target model, we must have prior knowledge of the phase transitions to generate the training data. This aspect is problematic when we aim at exploring new physics and unknown phenomena in the model. Feature engineering can also be problematic for this purpose because it requires prior knowledge of features that characterize the model or phase transitions. 

In this sense, the two methods discussed in this article, i.e., the PC method and the TI method, have advantages over previous methods. First, both the discussed methods require no or only minimal prior knowledge about properties of a target model such as transition temperatures, order parameters, and the number of phases. Specifically, only the number of phases must be known in advance for the PC method, while any of these properties do not have to be known for the TI method to detect the BKT transitions. Second, both the discussed methods require no or only minimal preprocessing of data for feature engineering to prepare the input data. It has been demonstrated that to detect the BKT transitions, only raw spin configurations without any preprocessing are required for the $q$-state clock model, while only the vortex configurations made from the spin configurations are required for the XXZ model. Note that some of the previous methods use vortex configurations, histograms of spin orientations, and spin correlation functions for input data, which are made from spin configurations via the feature engineering process. Among these, the vortex configuration may be the most natural feature for BKT transitions associated with binding and unbinding of vortices and antivortices. This quantity can be calculated from the spin-configuration data very easily.

\subsection{Comparison with the Monte Carlo methods}
The machine learning methods proposed in this article have the following two advantages over the Monte Carlo methods. First, they require much less computational cost. In the Monte Carlo methods, a large number of state-updates must be done to bring the system to thermal equilibrium. Furthermore, a large number of samplings is also required to accurately calculate the thermal averages of physical quantities. On the contrary, the machine learning methods do not require such heavy computational procedures. Certainly, it is necessary to generate spin configurations to prepare the input data using the Monte Carlo thermalization technique. However, they are not required to reach real thermal equilibrium in contrast to the Monte Carlo methods, but spin configurations on the way to the thermal equilibrium can also work as input data because they already contain information and features of phases and phase transitions. This enables us to reduce the computational time significantly. In addition, the training of neural network, which dominates the computational time of machine learning methods, takes typically one hour at most. This time scale is much shorter than that of the Monte Carlo methods, which is typically a few ten hours or even a few days. 

Second, the machine learning methods has advantages with respect to the scalability and generalizability. In the Monte Carlo methods, several physical quantities are calculated to determine the transition temperatures. Some of the physical quantities do not have general formula to be used for the Monte Carlo calculations, and we need to derive its specific expression for the target model. The helicity modulus used to detect the BKT transition in continuous spin systems is a typical example of such quantities. In such cases, it is time-consuming to derive the expression for each model. The derivation of its model-dependent formula is very difficult especially for complex spin models for real materials which contain magnetic anisotropies, further-neighbor exchange interactions and/or higher-order interactions. This problem hinders systematic research and exploration of phase transitions in various materials described by different spin models. The methods discussed in this article, on the other hand, do not require such derivation procedures. Once the neural network is trained by spin or vortex configurations of a well-known model, it can be applied to other spin models with excellent scalability and generalizability.

On the other hand, there are some disadvantages. First, it is difficult to systematically improve accuracy of the evaluations. We have demonstrated that both the PC and TI methods can obtain the transition temperatures in good agreement with those obtained by the Monte Carlo method. However, it is nontrivial how to make the value as close as possible to the true values. In the Monte Carlo methods, the statistical errors can be suppressed by increasing the number of samplings. On the contrary, it is not the case for the machine learning methods. Systematic methodology to improve accuracy has not been established at present. This is because there are many hyperparameters intricately entangled with each other in neural networks such as the number of training data, the number of training steps, the number of hidden layers, the number of nodes in each layer, choices of the activation functions etc.

Second, it is difficult to discern phase transitions from a heat map even if it is analyzed with the correlation function $C_W(T)$ and the variance $V_W(T)$. We have demonstrated that their temperature profiles exhibit apparent changes in slope at phase transition points for both the conventional second-order phase transitions and the topological BKT transitions. These profiles sometimes exhibit changes in slope also at other points which are not relevant to phase transitions directly. They are rather critical in the case of the BKT transition in the XY-like XXZ model with $\Delta$=0.95. Although further analysis identified points related with the temperature evolution of vortex density, we could not determine the BKT-transition point uniquely from the simple heat-map analysis only. In the TI methods, more advanced analysis may sometimes be required to determine phase-transition points. 

As discussed above, the PC method and the TI method have both advantages and disadvantages over the Monte Carlo method as a conventional method based on statistical mechanics. Currently, the machine learning method and the Monte Carlo method are complementary with each other. To realize the fully machine-learning-based research of physics in the future, it is necessary to clarify what the neural network sees and what the machine learning can do through repeating trials and accumulating experiences.

\section{Conclusion}
In this article, we have discussed recently proposed machine learning methods named PC method and TI method for detection of the BKT transitions in some classical spin models. The BKT transition is one of the typical topological phase transitions, which is not accompanied by spontaneous symmetry breaking in contrast to conventional phase transitions within Landau's scheme. The BKT transition is difficult to detect not only by conventional methods based on statistical mechanics such as the Monte Carlo method but also by simple machine learning methods. The PC method is a method to detect phase transitions in a target model by using a neural network trained with spin or vortex configurations of other well-known models so as to classify the phases correctly. On the other hand, the TI method is a method to detect phase transitions by analyzing the weight matrix connecting the last hidden layer and the output layer of the optimized neural network trained so as to correctly identify a temperature at which a given spin or vortex configuration is generated by the Monte Carlo thermalization technique. It has been demonstrated that these methods can detect both the second-order phase transitions and the BKT transitions successfully. 

Several machine learning methods for detection of the BKT transitions have been proposed and their demonstrations have been reported so far, but most of them are based on supervised learning techniques. The two methods discussed in this article have several advantages over them. One advantage is that the methods require no or less information of the model in advance. Another advantage is the least requirement of the feature engineering. These characteristics are suitable for systematic exploration of novel physics and phenomena in unknown spin models. The era that machine learning techniques are used for research of physics in earnest is approaching. We hope this article can contribute, even partially, to the development of this research field.

\section{Acknowledgment}
We are grateful to Yusuke Murata and Yasuhiro Tanaka for useful discussions. This work was supported by Japan Society for the Promotion of Science KAKENHI (Grants No.JP20H00337, No.JP23H04522 and No.JP24H02231), CREST, the Japan Science and Technology Agency (Grant No. JPMJCR20T1), Waseda University Grant for Special Research Projects (2023E-026, 2023C-140, 2024C-153, and 2024C-155), and Waseda Research Institute for Science and Engineering, Grant-in-Aid for Young Scientists (Early Bird).
%%%%%%%%%% References %%%%%%%%%%%%%%%%%%%

\end{document}